# THE INFLUENCES OF MAGNETIC SHEAR ON THE IMPROVEMENT OF THE QUALITY OF CONFINEMENT IN THE PLASMA OF TOKAMAK

M. El Mouden[1], D. Saifaoui[1], A. Dezairi[2], H. Imzi[1];
[1]Laboratory of Theoretical Physics, Faculty of Sciences- Ain Chock, Casablanca, Morocco,
[2]Laboratory of the Physics of Condensed Matter, Faculty of Sciences- Ben M'sik, Casablanca, Morocco,

**ABSTRACT:** In this paper we have studied the influence of reversed shear on the improvement of the confinement's quality in the plasma of tokamak and especially in reducing the anomalous transport. For that, we have used a special model for the drift wave fields. Comparison between particles trajectories for normal and reversed shear is carried out in 2 D and 3 D presentation. Also, the diffusion coefficient of particles for the two cases, normal and negative shear, is evaluated.

**INTRODUCTION:** The anomalous transport in the plasmas of tokamak has been studied for many years, all over the world, in order to control thermonuclear fusion. It is very well known that the destruction of the magnetic surfaces which are responsible for plasma confinement, the stochastic region's formation, the anomalous transport of particles and energy, etc, resulting in magnetohydrodynamic instabilities, electrostatic and magnetic turbulence and the phenomena of drift, drive to the deterioration of the confinement. Concerning this fact, all effort in this sense requires the understanding of the origin of this transport, of the instabilities that exist in the plasmas of the tokamak and the development of reliable theoretical models that are capable of describing the complex dynamics of fusion's plasma [3,5,6,7,8,9]. In this work, we are interested- firstly- in the survey of the dynamics of the lines of magnetic fields, and of plasma 's particles in the ulterior stage.

The other very important point that we will develop in this work consists of studying the contribution of the reversed shear to the improvement of the plasma's confinement in the tokamak and other rules of improvement of the confinement. Indeed, the research concentrates currently more on the control of these improved confinement régimes that have been observed during the last decade. These are the régimes where we observe the plasma diffusion's reduction, and what is impressive is the formation of the transport barrier. We say the word "barrier" because it opposes the motion of the particles anomalous diffusion taking the shape of an obstacle to particles of plasma. We studied the motion of the guiding center of particles, under the action of electrostatic field perturbation, and in a configuration known as of reversed magnetic shear. This configuration is adopted more and more in the new machines of fusion, since we observe important reductions there in the diffusion coefficients. The equations of the motion of the guiding center will be replaced by equations of the mapping known as Nontwist Standard Mapping, and in the phases space of the particles, we simulated the trajectory in two completely different cases of the safety factor profile: the normal profile and the reversed one. In the normal case, the trajectory can be described by the KAM theory, while in the reversed case the dynamic is described by the formation of the transport barrier of the particles that are localized in the neighborhood of the resonance surface that corresponds to the minimum of q. This surface confines plasma and reduces its diffusion.

## I – MAPPING EQUATIONS:

### I-1 Equation of motion:

In our survey, we use a simplified model of an equilibrium magnetic field, according to toroidal geometry, which is described by the following relation:

$$\vec{B} = B_\theta(r)\vec{e}_\theta + B_\varphi \vec{e}_\varphi \qquad (1)$$

consists of the poloïdal magnetic field component and the toroidal magnetic field component which are bound through the relation: $B_\theta(r) = \dfrac{r}{q(r)R_0} B_\varphi$, of which r is the minor radius of plasma, $\theta$ and $\varphi$ are respectively the poloïdal and toroidal angle, and finally, $q(r)$ is the safety factor.

In the Gauss units system, the equation of the motion of the guiding center is given by:

$$\frac{d\vec{x}}{dt} = v_{//} \frac{\vec{B}}{\|\vec{B}\|} + c \frac{\vec{E} \wedge \vec{B}}{B^2} \qquad (2)$$

where, $v_{//}$ is the parallel velocity, $\vec{E}$ and $\vec{B}$ are the electric and magnetic fields, and the last term of this equation represents the drift velocity.

The electric field satisfies the relation:
$$\vec{E} = -\vec{\nabla}\phi \qquad (3)$$

The correspondent electrostatic potential $\phi$ can be written as the sum of two terms, the first is the radial part supposed at equilibrium, and the second one represents the fluctuating part, noted $\tilde{\phi}$. We use the model of the spectrum of drift wave, and we have:

$$\tilde{\phi}(\vec{x},t) = \sum_{m,l,n} \phi_{m,l,n} \cos(m\theta - l\varphi - n\omega_0 t) \qquad (4)$$

where, $\omega_0$ is the lowest angular frequency in the spectrum of drift wave, and $\theta$ and $\varphi$ are random variables.

The disrupted electrostatic field $\tilde{E}$ is joined to the potential of perturbation with[10]: $\tilde{\vec{E}} = -\vec{\nabla}\tilde{\phi}$.

In the continuation of our survey, we suppose that $B \approx B_\varphi \gg B_\theta$ and $B_r = 0$ respect the system of toroidal coordinates $(r, \theta, \varphi)$ and we introduce $\overline{E}_r$ as the equilibrium radial electric field. The previous equation of motion will be projected in this system of coordinates, and we get the following system:

$$\frac{dr}{dt} = -\frac{c}{B}\frac{1}{r}\frac{\partial \tilde{\phi}}{\partial \theta}$$

$$r\frac{d\theta}{dt} = v_{//}\frac{B_\theta}{B} + \frac{c}{B}\frac{\partial \tilde{\phi}}{\partial r} - \frac{c\tilde{E}_r}{B} \quad (5)$$

$$R\frac{d\varphi}{dt} = v_{//}$$

Substituting (4) into (5), we obtain:

$$\frac{dr}{dt} = -\frac{c}{Br}\frac{\partial}{\partial \theta}\sum_{M,L,n}\phi_{mM,L}\begin{bmatrix}\cos(M\theta - L\varphi)\cos(n\omega_0 t)\\+\sin(M\theta - L\varphi)\sin(n\omega_0 t)\end{bmatrix} \quad (6)$$

Using two important properties of functions sin and cos which are:

$$\sum_{n=-\infty}^{+\infty}\sin(n\omega_0 t) = 0 \quad \text{and}$$

$$\sum_{n=-\infty}^{+\infty}\cos(n\omega_0 t) = 2\pi\sum\delta(\omega_0 t - 2\pi n)$$

Then we have:

$$\frac{dr}{dt} = \frac{2\pi c}{Br}\sum_{M,L}M\phi_{M,L}\sin(M\theta - L\varphi)\delta(\omega_0 t - 2\pi n) \quad (7)$$

Hence this model spectrum gives impulsive jumps in $r$ at time $t_n = \frac{2\pi n}{\omega_0}$.

### I-2 Transformation into a mapping equation:

It is more useful to replace the equations of the guiding center particles motion with those of approach mapping. For that, we introduce some new canonical variables.
We define these new angle-action variables $(\chi, J)$ as:

$$J = \left(\frac{r}{a}\right)^2 \text{ and } \chi = M\theta - L\varphi \quad (8)$$

$a$ is the minor radius of the torus. Also we suppose that only one perturbation mode $(M,L)$ dominates in the equation of evolution of $r$. From equation (7) we have:

$$\frac{dJ}{dt} = \frac{2r}{a^2}\frac{dr}{dt} = \frac{4\pi c}{a^2 B}M\phi_{M,L}\sin(M\theta - L\varphi)\times\sum_n\delta(\omega_0 t - 2\pi n) \quad (9)$$

By integration over one jump at time $t_n = \frac{2\pi n}{\omega_0}$, equation (5) becomes in terms of $\chi$:

$$\frac{d\chi}{dt} = M\frac{B_\theta}{rB}(v_{//} - \frac{c\overline{E}_r}{B_\theta}) - L\frac{v_{//}}{R} \quad (10)$$

Ignoring in first approximation of $\overline{E}_r$, the integration of the $\chi$ and $J$ differential equations as the function of time, then, we get their evolution equation:

$$J_{N+1} = J_N + \frac{4\pi c}{a^2 B_0}\frac{M\phi}{\omega_0}\sin(M\theta_N - L\varphi_N) \quad (11)$$

$$\chi_{N+1} = \chi_N + \frac{2\pi}{\omega_0}\frac{v_{//}}{qR}(M - Lq) \quad (12)$$

### I-3 The Standard Nontwist Mapping (SNM):

The important parameter that is going to allow us to construct this model is the safety factor, that is given as it has already been specified, by the relation $q(r) = \frac{rB_\varphi}{RB_\theta}$.

We suppose that this safety factor has a minimum local in the neighborhood of a certain value $r_m$, it means that $q_m = q(r_m); q'(r_m) = 0$, since

$$\left.\frac{dq}{dJ}\right|_{r=r_m} = \left.\frac{dq}{dr}\right|_{r=r_m}\left.\frac{dr}{dJ}\right|_{r=r_m} = 0.$$

Then $q$ possesses a minimum at $J_m = J(r_m)$.

We are interested in the motion of particles in the neighborhood of $r_m$, and we do a Taylor development of $q$ around $J_m$, we can write such development as:

$$q(J) = q(J_m) + \frac{q_m''}{2}(J - J_m)^2$$

While substituting this last equation in (10), we get:

$$\frac{d\chi}{dt} = \frac{v_{//}}{Rq_m}\left(M - Lq_m - \frac{Mq_m''}{2q_m}(J - J_m)^2\right) \quad (13)$$

and after integration on the step of time $\frac{1}{\Delta\omega}$ we obtain:

$$\chi_{N+1} = \chi_N + \frac{2\pi}{w_0}\frac{v_{//}}{Rq_m}\left(\delta - \frac{Mq_m''}{2q_m}(J_{N+1} - J_m)^2\right) \quad (14)$$

where $\delta = M - Lq_m$

We introduce the dimensionless variables K and T such that:

$$K = \frac{\chi}{2\pi}; \quad T = \left(\frac{Mq_m''}{2q_m\delta}\right)^{1/2}(J - J_m) = k(J - J_m) \quad \text{Hence,}$$

we can transform the system of mapping equation in the form of the SNM:

$$K_{N+1} = K_N + \frac{v_{//}\delta}{Rq_m\omega_0}(1 - T_{N+1}^2) = \chi_N + \alpha(1 - T_{N+1}^2)$$

$$T_{N+1} = T_N + \left(\frac{2\pi c M\phi}{a^2 B\omega_0}\right)\left(\frac{2Mq_m''}{q_m\delta}\right)^{1/2}\sin(2\pi K_N) \quad (15)$$

$$T_{N+1} = T_N - \beta\sin(2\pi K_N)$$

where $\alpha = \frac{v_{//}\delta}{Rq_m\omega_0}$;

$$\beta = \left(\frac{-2\pi c M\phi}{a_2 B\omega_0}\right)\left(\frac{2Mq_m''}{q_m\delta}\right)^{1/2}$$

### I-4 Global Mapping:

In order to introduce the effects of reversed magnetic shear and the radial electric field, we define the

system of equation of global mapping that is the following:[7,11,10]

$$J_{N+1} = J_N + \frac{4\pi c}{a^2 B_0} \frac{M\phi}{\omega_0} \sin(M\theta_N - L\varphi_N) \quad (16)$$

$$K_{N+1} = K_N + RK1(J_{N+1}) + RK2(J_{N+1}) \quad (17)$$

Where:
$$RK1(J) = \frac{v_{//}(J)}{\omega_0 q R}(M - Lq(J)) \quad (18)$$

$$RK2(J) = -\frac{cM}{\omega_0 a B_0} \frac{\overline{E}_r(J)}{\sqrt{J}} \quad (19)$$

$$v_{//}(J) = \sqrt{\frac{2}{m}(\zeta_t - e\Phi_0(J))(1 - \lambda B_0)} \quad (20)$$

with: $K = \frac{\chi}{2\pi}$, $\zeta_t$: the total initial energy, $e$: the particle charge, $\lambda = \mu/\zeta_t$ where $\mu$ is the magnetic moment and $\Phi_0$ is the equilibrium potential as

$$\overline{E}_r(J) = -\frac{\partial \Phi_0}{\partial r}\bigg|_{r=a\sqrt{J}}.$$

**Choice of the profiles of the factor of security :**

The trajectories will already be solved numerically according to this formulation of established mapping. However, we are going to return more importance to the control parameter $q$, and we are going to work in two configuration cases that are completely different.

We define two $q$-profiles of (**Figure 1.**):

℘ Normal: $q(r) = 1.99 + 1.94\left(\frac{r}{a}\right)^2$    (21)

℘ Reversed: $q(r) = 1.99 + 7.76\left(\frac{r}{a} - 0.5\right)^2$    (22)

The choice of the potential $\Phi_0$ depends on the nature of the $q$-profile, therefore, in the normal case we use $\Phi_0(r) = -\Phi_0\left(1 - \left(\frac{r}{a}\right)^2\right)$ and for the reversed profile we take $\Phi_0(r) = -\Phi_0\left(1 - \left(\frac{r}{a}\right)^2\right)\exp(1 - r/a)$.

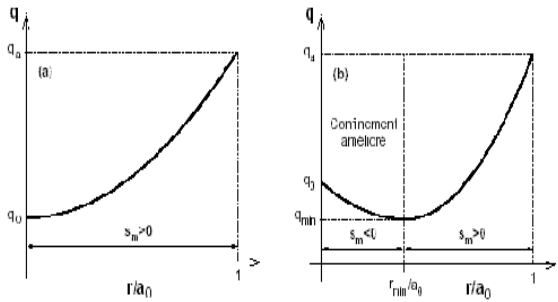

**Figure 1.** Profile of the safety factor versus r/a for reversed (right hand) and normal (left hand) shear.

## II - NUMERICAL INTEGRATION AND INTERPRETATIONS :

In the next section, we neglect the electric field radial component and investigate the map phase structure by calculating 1000 massive $D^+$ trajectories with various initial conditions in configuration spaces for the reversed and normal shear cases of the safety factor.

In the calculation below, we have used the Texas Experimental Tokamak (TEXT) system parameters, with major radius $R_0 = 100cm$, minor radius $a = 26cm$, and center-line field $B = 3tesla$.

We took; $\omega_0 = 1.93 * 10^5$; and we chose the mode of perturbation (M=12, L=6) and $\lambda = \frac{\mu}{\zeta}$ with $\zeta = 167\ eV$: the energy of the particles.

**Simulations & interpretations :**

In presence of the electrostatic perturbations and the normal profile of the safety coefficient $q$, the stochasticity of the trajectories increases and it is the principle reason for the particles' diffusion through the magnetic surfaces (**Figure 2.** (b)).

While in the case of reversed shear, the principle result shows that the transport barrier is near of surface that corresponds to the minimal value of $q$ exists. This barrier plays an important role on the reduction of the transport and the diffusion of the particles, which drives to the improvement of the plasma confinement (**Figure 2.** (a)).

In the case of the 3- dimensional simulation, we observe the same phenomena as those in the case of 2-dimensional simulation except that we do not see the formation of islands, and what we observe is a transition of the particles from the regions ($r/a < 0.5$) toward regions ($r/a > 0.5$).

While increasing the amplitude of the perturbation, and in the reversed case (**Figure 3.** (a)) we observe that this transition drives to a transition of this barrier toward the outside regions of the tokamak ($r/a \to 1$) to prevent the diffusion of these particles. Whereas in the normal case (**Figure 3.** (b)), we observe that the majority of the particles escaped which means that we observe the total destruction of the magnetic surfaces of the confinement.

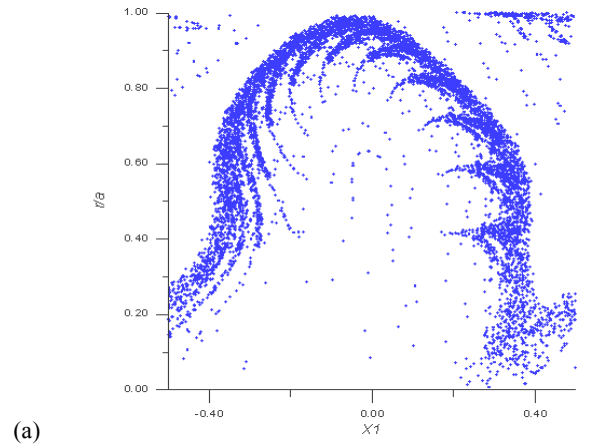

(a)

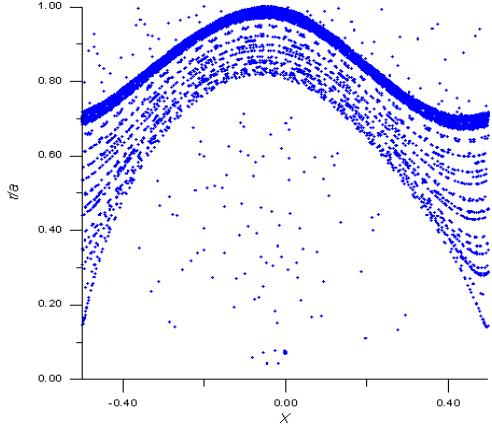

(b)

**Figure 2.** Poincaré Section for 1000 particles in the $(\chi, r/a)$ plane for the value 1.5 $eV$ of the perturbation: (a) reversed shear ; (b) normal shear.

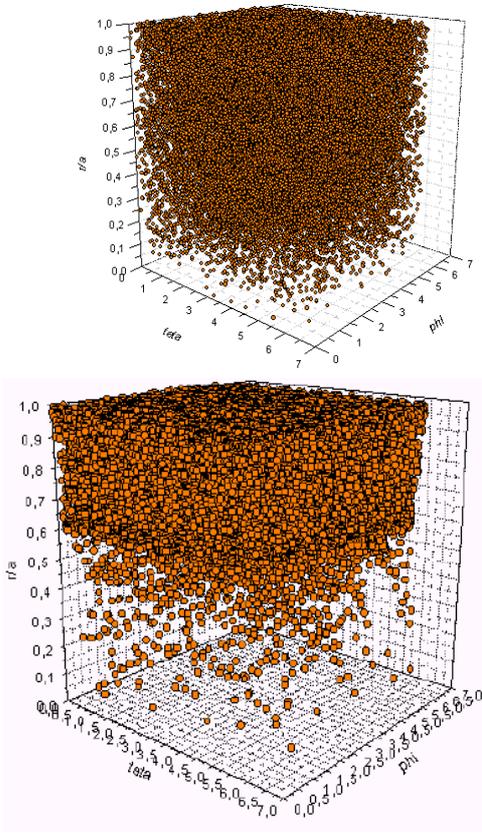

(b)

**Figure 3.** Simulation of the trajectories of 1000 particles at 3- dimensional $(\theta, \varphi, r/a)$ for the value 1.1 $eV$ of the perturbation: (a) reversed shear ; (b) normal shear.

## III - THE DIFFUSION OF THE PARTICLES :

To evaluate the plasma particles' diffusion through the magnetic surfaces within the tokamak reactors, we simulated in two case the particulate diffusion coefficients. We expressed it under the most classic formula as follows:

$$D = \lim_{N \to \infty} \frac{\langle (r_N - r_0)^2 \rangle}{2 t_N} \quad (23)$$

and we represented time evolution of the ratio of diffusion coefficient in reversed shear (*Dr*) on diffusion coefficient in the normal case (*Dn*) for deferent values of the perturbation (**Figure 4.**). Here *tn* is a time step, We observe a large reduction of particles diffusion in the reversed case due to the transport barrier which have tendency to suppress the anomalous transport. The diffusion in the reversed case become lower than those in the normal case. And the formation of transport barrier which suppress turbulent transport may explain this reduction [4].

Finally, we represented the same ratio but now as function of the amplitude of perturbation for deferent values of *tn* (**Figure 5.**).

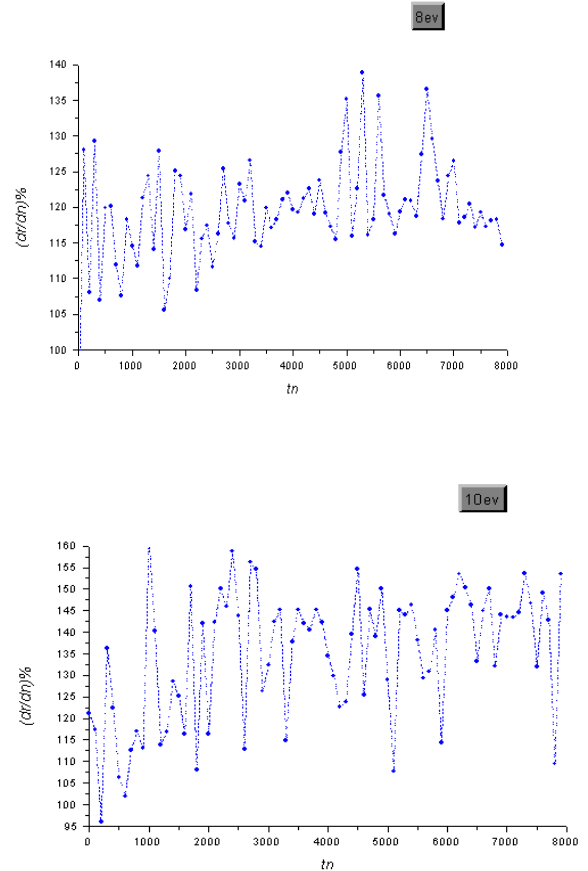

**Figure 4.** The ratio between the diffusion coefficient in reversed and normal shear as function of *tn*.

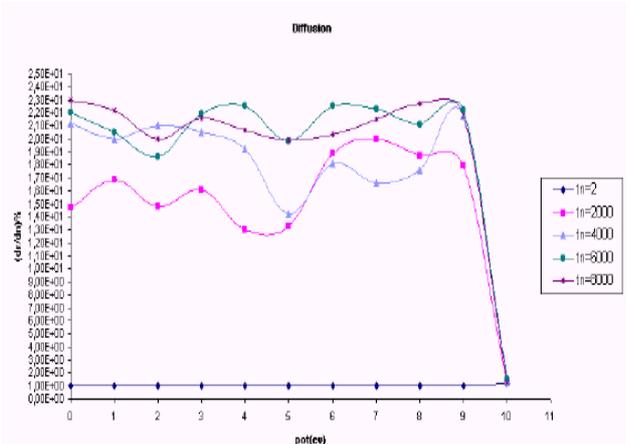

**Figure 5.** The ratio between the diffusion coefficient in reversed and normal shear as function of the amplitude of perturbation.

## IV- RADIAL ELECTRIC FIELD:

(a)

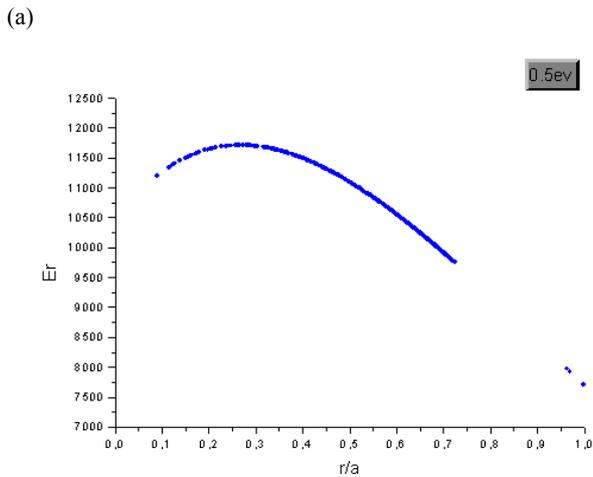

(b)

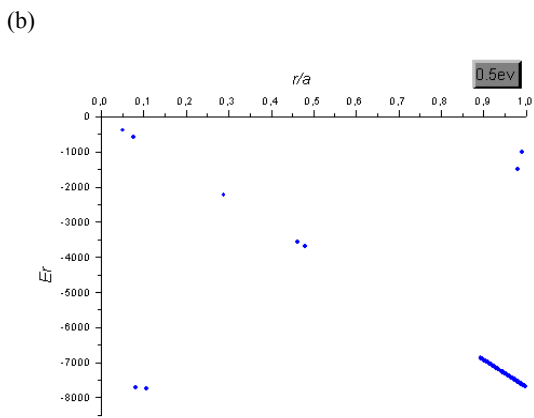

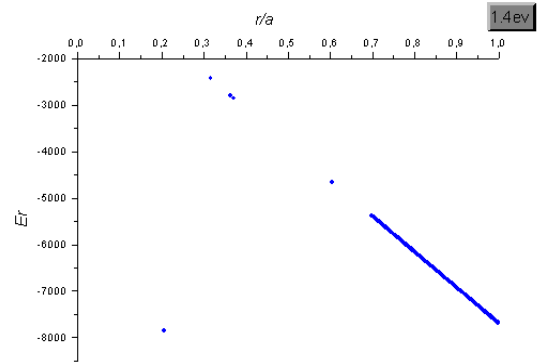

**Figure 6.** Radial electric field in as function of *r/a* for the value 0.5 *eV* and 1.5 *eV* of the perturbation in reversed and normal shear: (a) reversed shear ; (b) normal shear.

At the neighbourhood of the plasma border, the radial electric field *Er* decreases, what drives to the reduction of the potential electrostatic *Fr*, thereafter to the reduction of the flight of the plasma ions and therefore the reduction of the anomalous transport with the shear.

## CONCLUSION & PERSPECTIVES:

Anomalous transport observed in tokamaks is known as the result of the electrostatic and magnetic turbulence. thus, in the presence of electric perturbation and for the normal profile of the safety factor *q*, the stochasticity of the trajectories increases and this is the principal cause of diffusion of particles through magnetic surfaces. However for the reversed shear case, the most important result is the impressive formation of a strong transport barrier, which is localized near of minimum value of *q* (*q* is the safety factor) (**Figure 2.** (a)). This barrier plays a very important role in the improvement of the plasma confinement while preventing its radial diffusion. To evaluate quantitatively the diffusion, we simulated from the Mapping equations, the diffusion coefficient in each of the two previous cases, and we drew the ratio that shows a clean reduction in the diffusion observed in the reversed magnetic shear profile. Therefore, the diffusion decreases, the confinement improves and the control of the fusion reactors to function in these modes permits the reduction of the anomalous transport in the tokamaks.

Lately, there has been consensus on an international scale to construct a big experimental reactor in Cadarache ITER: International Thermonuclear Experimental Reactor. In this project, Europe, the USA, Canada, Australia, Japan and China will participate. The launch of this project in France is going to be a big step in the success of thermonuclear energy production. In this setting, and in perspective, we are going to reproduce the same simulations with the parameters of ITER to be able to compare them with those obtained for TEXT.

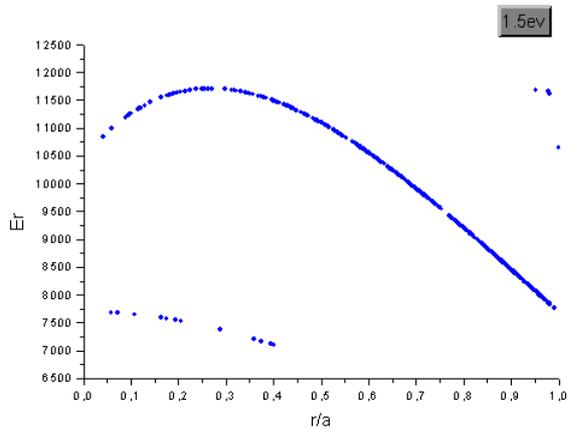